# Correlated and Multi-frequency Diffusion Modeling for Highly Under-sampled MRI Reconstruction

Yu Guan, Chuanming Yu, Shiyu Lu, Zhuoxu Cui, Dong Liang, *Senior Member, IEEE*,
Qiegen Liu, *Senior Member, IEEE*

*Abstract*—Most existing MRI reconstruction methods perform targeted reconstruction of the entire MR image without taking specific tissue regions into consideration. This may fail to emphasize the reconstruction accuracy on important tissues for diagnosis. In this study, leveraging a combination of the properties of k-space data and the diffusion process, our novel scheme focuses on mining the multi-frequency prior with different strategies to preserve fine texture details in the reconstructed image. In addition, a diffusion process can converge more quickly if its target distribution closely resembles the noise distribution in the process. This can be accomplished through various high-frequency prior extractors. The finding further solidifies the effectiveness of the score-based generative model. On top of all the advantages, our method improves the accuracy of MRI reconstruction and accelerates sampling process. Experimental results verify that the proposed method successfully obtains more accurate reconstruction and outperforms state-of-the-art methods.

## I. INTRODUCTION

Magnetic Resonance Imaging (MRI) is an essential technology that employs the nuclear magnetization of the material for imaging without ionizing radiation [1], [2]. Despite its numerous advantages, the limited imaging speed of MRI remains a significant bottleneck. To address this issue, compressed sensing [3]-[6] and parallel imaging [7]-[9] have been developed. While advancements in their reconstruction algorithms have led to significant progress in clinical settings, challenges regarding reconstruction accuracy persist. Therefore, optimizing reconstruction processes to capture as many intricate details as possible has become a primary area of focus in MRI research.

In recent years, diffusion models have gained wide interest as a new class of generative model since they can provide a more accurate representation of the data distribution and surprisingly high sample quality [10]-[17]. Hereafter, leveraging the learned score function as a prior, the diffusion model with Langevin dynamic appeared as an emerging technology to reconstruct MRI. Among many works, Quan *et al.* [16] presented a diffusion framework to exploit homotopic gradients of generative density priors by taking advantage of the denoising score matching for MRI reconstruction. A similar approach has been also applied to the accelerated MRI by Chung *et al.* [17], which trained a continuous time-dependent score function with denoising score matching for the high accuracy of reconstruction. Thanks to the development of diffusion model in MR reconstruction has shown impressive results, the hotspot of research has transitioned to explore the possibility of extracting prior information in the high-frequency space to improve the performance of accurate reconstruction [18]-[20].

Generally, there are two prominent strategies have garnered attention for the extraction of high-frequency information, denoted as "Mask-K-Space" and "Weight-K-Space". From a visual perspective, "Mask-K-Space" directly segregates the high-frequency and low-frequency components of k-space data through the application of artificial masks. Mathematically, this approach exhibits a resemblance to the concept of hard-thresholding. One of the most promising approaches was proposed by Xie *et al.* [21], they applied diffusion process in k-space domain with conditioned under-sampling mask and obtained the high-frequency information of acquired data through the imaging operator. Diverging from the hard-threshold technique employed by "Mask-K-Space", "Weight-K-Space" employs a weight-based technology to modulate the entirety of k-space data, which is similar to the underlying principle of soft-thresholding for data manipulation. Following this idea to improve the reconstruction accuracy of MRI has been successfully demonstrated in our earlier work [22]. In detail, we effectively applied k-space weight-based techniques in score-based generative model to capture high-frequency priors. Another common approach depending on the idea of "Weight-K-Space" was presented by Cao et al. [23], where the goal was to acquire high-frequency noise with soft-thresholding and combine it with image data to construct a new diffusion model for robust MRI reconstruction.

Nevertheless, there are still many open questions and challenges in this field. For example, whilst our initial weight-based strategy was found to be successful, it demonstrated a high level of dependence on specific parameter selection. Furthermore, the ability to improve the reconstruction accuracy by weight-based technology itself has proved to be limited, even with optimal parameters. Therefore, the utilization of supplementary manners to construct an array of varied and extensive priors may serve as an effective solution. Other shortcomings of the above methods were that they were not universal due to the measurement-conditioned model [21] and suffered from long computation time due to the transformation of different domains [23]. Based on the above analysis, we introduce a novel reconstruction scheme that inherits the advantages of previous methods while eliminating some of their shortcomings.

First of all, the aim of the study is to focus on the underlying signal properties of k-space data in high-frequency space

This work was supported in part by National Natural Science Foundation of China under 62122033 and Key Research and Development Program of Jiangxi Province under 20212BBE53001. (Corresponding authors: D. Liang and Q. Liu)

Y. Guan, C. Yu and S. Lu are with School of Mathematics and Computer Sciences, Nanchang University, Nanchang 330031, China. ({guanyu, yuchuanming, lushiyu}@email.ncu.edu.cn)

Q. Liu is with School of Information Engineering, Nanchang University, Nanchang 330031, China. (liuqiegen@ncu.edu.cn)

Z. Cui and D. Liang are with Research Center for Medical AI, Shenzhen Institutes of Advanced Technology, Chinese Academy of Sciences, Shenzhen 518055, China. ({zx.cui, dong.liang}@siat.ac.cn)

through the combination of "Weight-K-Space" and "Mask-K-Space". Afterward, a unique method named **Co**rrelated and **M**ulti-frequency **D**iffusion **M**odeling (**CM-DM**) is put forward to preserve high-frequency content as well as fine textural details in the reconstructed image. On one hand, integrating two distinct strategies for extracting high-frequency information may optimize the utilization of complementary information characteristics, which is superior to the usage of any one strategy in isolation. On the other hand, it is noteworthy that the target distribution of the high-frequency diffusion process bears a closer resemblance to the noise distribution than that of the diffusion process over the entire k-space. The main contributions and observations of this work are summarized as follows:

- *Diversity of High-frequency Priors for Diffusion Modeling.* Aiming at accurate reconstruction, multi-profile high-frequency components in k-space domain are combined to directly train diffusion model. Experimental results demonstrate that when subjected to a high acceleration rate of 15-fold, the preservation of image details is enhanced due to the diffusion process primarily focuses on high-frequency.
- *A New Look to the Interpretation of Optimal Diffusion Time.* Constraining the diffusion process in the frequency domain and exploiting the structural distributional priors in high-frequency is an underlying contributor for convergence. **Theorem 1** illustrates how the feature can promote fast convergence of the diffusion process.

The remainder of this paper is exhibited as follows: Section II briefly introduces some relevant works in this study. Section III contains the key idea of the proposed method. The experimental setting and results are shown in Section IV. Section V conducts a concise discussion and Section VI draws a conclusion for this work.

## II. RELATED WORK

### A. Forward Imaging Model

The forward model of MRI can be represented by the following formula:

$$f = Ak + \eta \quad (1)$$

where $f \in \mathbb{R}^n$ is the under-sampled measurement in the k-space domain, $k \in \mathbb{R}^n$ is the medical k-space data to be reconstructed, $A \in \mathbb{R}^{n \times n}$ is a measuring matrix according to $k$, and $\eta$ is the Gaussian noise. More specifically, $A = PS$ for the sake of multi-coil acquisition, where $P$ is the under-sampling operator and $S$ is the coil sensitivity. Due to insufficient information of $k$, reconstructing accurate k-space data $k$ is known as an ill-posed issue. This means that we cannot obtain a solution $k$ by directly inverting Eq. (1). Thereby, it is important to impose constraints to achieve regularization and Eq. (1) can be formulated as an optimization problem:

$$\underset{k}{Min} \|Ak - f\|_2^2 + \lambda R(k) \quad (2)$$

where $\|Ak - y\|_2^2$ term is the data fidelity. $R(k)$ is the regularization term, which plays a significant role in pursuing high-quality results. $\lambda$ is the weight coefficient that balances the data fidelity term and the regularization term for optimization solution.

### B. Diffusion Models in MRI Reconstruction

Diffusion models are a cutting-edge class of generative models that have demonstrated to be highly effective in learning complex data distributions, such as MRI reconstruction [24]-[26]. Specially, Jalal *et al.* [27] proposed the first study which trains the diffusion model on MR images as prior information for the inversion pathway in reconstructing realistic MR images. Furthermore, it inspired Song *et al.* [28] to improve the basic theory of the original diffusion model and then derive it to the application of medical image reconstruction. As another class of relevant diffusion model, Gungor *et al.* [29] leveraged an efficient adaptive diffusion prior trained via adversarial mapping over large reverse diffusion steps for accelerated MRI reconstruction. Overall, diffusion model has proven to be one of the highly flexible and tractable generative models that can accurately generate complex data distributions from random noise in the image domain.

Recently, the transformation of the diffusion process from the image domain to the k-space domain to directly process k-space data has become a new research hotspot. Tu *et al.* [22] trained an unsupervised diffusion model on weighted and high-dimensional k-space data, where weight-based technology and high-dimensional space augmentation design are applied to the initial k-space data for better capturing the prior distribution. Similar to this work, Xie *et al.* [23] presented a measurement-conditioned diffusion model for MRI reconstruction and achieved results that outperformed other methods. They defined the model in the k-space domain with conditioned under-sampling mask to provide an estimate of uncertainty as output. Therefore, constraining the diffusion process in k-space domain not only enables the direct utilization of inherent k-space data for enhanced efficiency but also achieves substantial results.

With the remarkable advancements in above-mentioned methods, achieving more refined MRI reconstruction by extracting a set of high-frequency components has also observed a growing interest [30]. For instance, He *et al.* [31] extracted high-frequency part of images and designed a multi-profile denoising autoencoder for attaining deep frequency-recurrent prior. Besides, Yang *et al.* [32] developed a two-stage reconstruction procedure to treat the low-frequency and high-frequency parts progressively for a more accurate reconstruction. Influenced by the diffusion model, Xie *et al.* [21] invented an effective algorithm measurement-conditioned denoising diffusion probabilistic model, which exploited high-frequency information in k-space domain with a specific mask. In the follow-up development, a score-based diffusion method is developed by Cao *et al.* [23] that they added high-frequency noise to the data in image domain for stable and accelerated MR reconstruction.

## III. METHOD

### A. Motivation

K-space data contains high-frequency and low-frequency components. The texture details of the image are associated with high-frequency components while the low-frequency components can effectively represent image profiles. Exploiting high-frequency component is significant for realizing accurate reconstruction. Considering that more attention is paid to handling high-frequency information for accurate MRI reconstruction. In this section, we first focus on different strategies of extracting high-frequency information in the k-space domain. Critically, it is shown that the convergence can be sped up dramatically when we constraint the diffusion process in the high-frequency domain, which is further briefly discussed in **Section III.B**. Then we exploit the strengths of each strategy and combine them in different ways to avoid the unique pitfalls of each (**Section III.C**). **Fig. 1** shows the overall procedure of the proposed method.

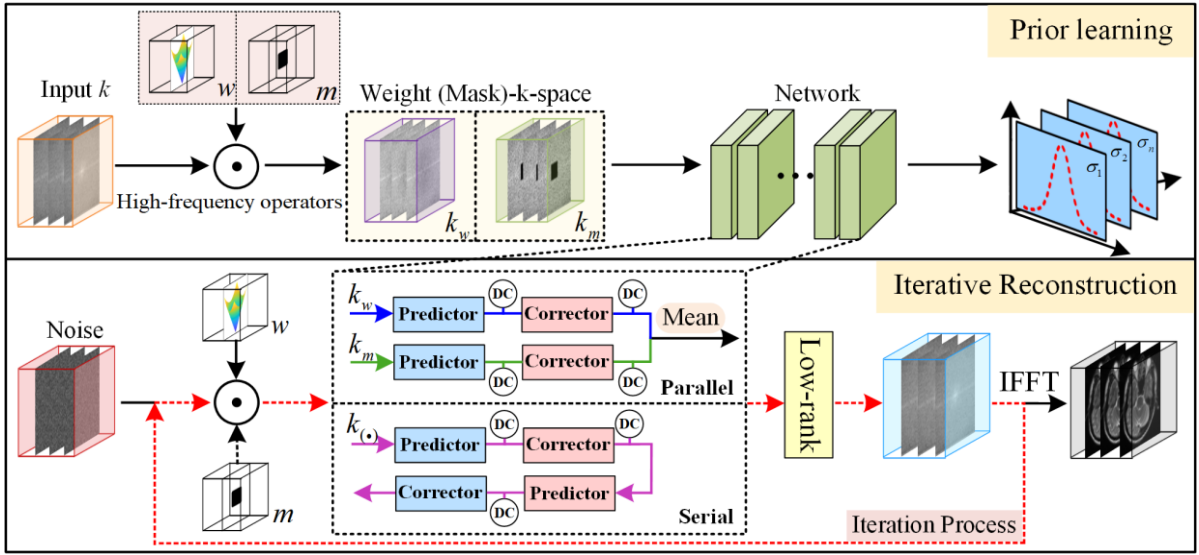

**Fig. 1.** Overview of the CM-DM method. Different high-frequency prior extractors ("Weight-K-Space" or "Mask-K-Space") are employed to restrict the diffusion process in the training process. The second row shows that the input goes through predictor and corrector iteratively to obtain the final reconstructed image.

### B. Diffusion Process Meets Multi-frequency Space

**Diffusion Process in K-Space:** Suppose a certain dataset $x_0 \in \mathbb{R}^d$ contains i.i.d. sampled data from an unknown distribution $x_0 \sim p_x(X)$ and the score function $s_\theta(x)$ with parameter $\theta$ is an approximation of the gradient of its log probability density, i.e., $\nabla_x \log p_x(X)$. Then, diffusion process sampling is performed according to the score function to obtain samples that obey the distribution $p_x(X)$, i.e.,

$$x^{t-1} = x^t + \frac{\varepsilon^t}{2}\nabla_x \log p_x(x^t) + \sqrt{\varepsilon^t} z^t, \ t=T,\cdots,1 \quad (3)$$

where $\varepsilon^t$ specifies the step size. The initial distribution $x^T$ is sampled from a given prior distribution and the noise $z^t$ is samples of the standard $d$-dimensional Gaussian distribution. Typically, the step size $\varepsilon^t$ for sampling must be small enough for ensuring stability in the diffusion procedure. Otherwise, the model will fail to converge to the target distribution [33]. As a result, the diffusion process suffers from a lengthy sampling process as many iterations are needed. Based on the theory in [34] that this limitation can be alleviated by increasing the similarity between the target data distribution and diffusion related distributions. This is intuitive as both the starting point and added noise is constrained to be relevant w.r.t. the target distribution, hence minimizing the inefficient fluctuations over the diffusion process. We first show that the sampling process can be equivalently transformed to the frequency domain as follows:

$$k^{t-1} = k^t + \frac{\varepsilon^t}{2}\nabla_k \log p_k(k^t) + \sqrt{\varepsilon^t} F[z^t] \quad (4)$$

where $F$ is Fourier transform and $k = Fx$. Given the same sampling process as **Eq. (1)**, the above formula aims to redesign the diffusion process for model inference in frequency domain that it is still capable of producing samples of excellent quality under high acceleration. Details are given in **Appendix A.1** for a full definition.

Nevertheless, there exists a general observation that the amplitude of low-frequency information of images is typically much higher than that of high-frequency ones [35]. This means the distribution of images exhibits huge gaps in quantity between different coordinates in the frequency domain, causing a severe ill-conditioned issue. Therefore, it explains the necessity to regulate the frequency distribution of a diffusion process, which is implemented by extracting only high-frequency information. With the above theoretical findings, as an instantiation we formulate different high-frequency operations to regulate the frequency distribution of the samples and promote the behavior of accelerated diffusion process. Concretely, the key idea is that mathematically matrix operator is effective in excavating high-frequency information in a way that the amplitude in frequency domains become more consistent along all the directions.

**Weight-K-Space:** To restrict the magnitude difference of data in k-space, the weight-based matrix strategy is presented to solve the problem that the value range of k-space data varies drastically. Specifically, it is incorporated to handle k-space data with high-frequency and low-frequency information during the training stage. The weight-based matrix operator can be expressed as follows:

$$K_w(k) = w \odot k; \ w = (r \cdot x_k^2 + r \cdot y_k^2)^p \quad (5)$$

where $k$ denotes the initial data in the k-space domain, and $w$ is the specific weight-based matrix. The $\odot$ means element-wise multiplication and $r$ is introduced for setting the cutoff value. $p$ decides the smoothness of the weight boundary while $x_k$ and $y_k$ are the count of frequency encoding lines and phase encoding lines. Formally, we enrich the above Langevin dynamics (**Eq. (4)**) by imposing a weighting operation into the diffusion process as:

$$K_w^{t-1} = K_w^t + \frac{\varepsilon^t}{2}\nabla_{K_w}\log p_{K_w}(K_w^t) + \sqrt{\varepsilon^t}F[z^t] \quad (6)$$

In most cases, it is weighted on the overall k-space data which has the advantage of strong convergence and theoretical analysis. However, the training of the method for MRI reconstruction problem is not easy in the presence of large number of parameters, which leads to the difficulty in fine-tuning parameters. For more fine-grained regulation, we can design multiple manners each with a dedicated advantage.

*Mask-K-Space:* Focusing on the extraction of high-frequency information while keeping the steady-state distribution simultaneously, another scheme based on k-space segmentation is proposed to obtain high-frequency information by adjusting window size. In detail, for the sake of effectively narrowing the range of values in the k-space data, the "Mask-K-Space" with an adjustable central window is designed to eliminate low-frequency information. Algorithmically, the specific "Mask-K-Space" operator is defined as follows:

$$K_m(k) = (I - m) \odot k \quad (7)$$

where $m$ is the specific "Mask-K-Space" kernel to separate low-frequency and high-frequency information, with value 1 in k-space center (assume the window size $n \times n$ is adjustable) and 0 for the rest of the region which denotes the high-frequency part. $I$ is an all-ones matrix and $(I - m)$ is a mask which could remove the low-frequency information in the central area. Combining the operation above, we redefine another diffusion process as:

$$K_m^{t-1} = K_m^t + \frac{\varepsilon^t}{2}\nabla_{K_m}\log p_{K_m}(K_m^t) + \sqrt{\varepsilon^t}F[z^t] \quad (8)$$

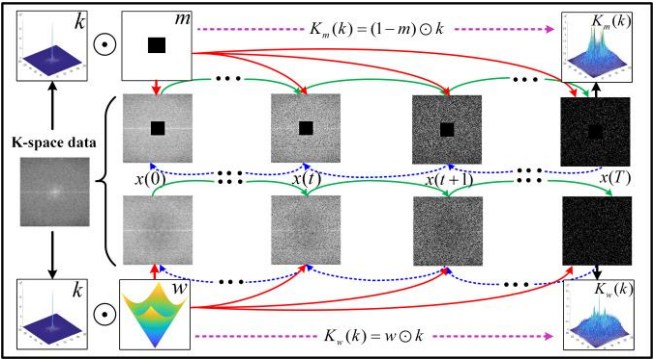

**Fig. 2.** Illustration of the diffusion process in high-frequency space. The first row shows the diffusion process of operator "Mask-K-Space" and the second row shows the diffusion process of operator "Weight-K-Space". Note that the amplitude of the k-space data becomes more uniform through different frequency domain operators.

*Deviation Analysis:* For theory completeness, we further briefly discuss the possibility to construct preconditioning operators using the matrix $w$ and $m$ (**Eq. 5 and Eq. 7**) as accelerators of the diffusion process. Note, this is a theoretical extension as formulated above to clarify the effect of high-frequency information on convergence (see **Fig. 2**).

**Theorem 1.** Suppose the diffusion processes (**Eq. (6)** and **Eq. (8)**) converge to a distribution $K_{(\cdot)}^* \sim p_{k^*}(K_{(\cdot)})$. Denoting $\mathcal{F}_{-t}$ as the $\sigma$-algebra generated by $\{K_{(\cdot)}^T, z^s, s = T, \cdots, t+1\}$. If the additive noise $\{z^s\}_{s=t+1}^T$ depend on $K_{(\cdot)}^t$ and satisfy the condition $\mathbb{E}[z^t|\mathcal{F}_{-t}] = 0, \forall t \in \{T, \cdots, 1\}$, then the deviation of $K_{(\cdot)}^t$ from $K_{(\cdot)}^*$ can be written as

$$\mathbb{E}[\|K_{(\cdot)}^* - K_{(\cdot)}^{t-1}\|^2] = C_1 + \varepsilon^t\mathbb{E}[\|z^t\|^2] - 2\sqrt{\varepsilon^t}\mathbb{E}[K_{(\cdot)}^* \cdot z^t] \quad (9)$$

where the term $C_1$ is a constant independent of $z^t$. $K_w^t$ and $K_m^t$ are an instantiation of $K_{(\cdot)}^t$.

***Proof.*** For the conditional deviation, we have

$$\begin{aligned}
&\mathbb{E}[\|K_{(\cdot)}^* - K_{(\cdot)}^{t-1}\|^2 \mid \mathcal{F}_{-t}] \\
&= \mathbb{E}[\|K_{(\cdot)}^* - K_{(\cdot)}^t - \frac{\varepsilon^t}{2}s_{K_{(\cdot)}^*}(K_{(\cdot)}^t) - \sqrt{\varepsilon^t}z^t\|^2 \mid \mathcal{F}_{-t}] \\
&= \mathbb{E}[\|K_{(\cdot)}^* - K_{(\cdot)}^t - \frac{\varepsilon^t}{2}s_{K_{(\cdot)}^*}(K_{(\cdot)}^t)\|^2 \mid \mathcal{F}_{-t}] + \varepsilon^t\mathbb{E}[\|z^t\|^2 \mid \mathcal{F}_{-t}] \\
&\quad - 2\sqrt{\varepsilon^t}\mathbb{E}[K_{(\cdot)}^* \cdot z^t \mid \mathcal{F}_{-t}]
\end{aligned} \quad (10)$$

The last equation is due to

$$\begin{aligned}
&\mathbb{E}[K_{(\cdot)}^t + \frac{\varepsilon^t}{2}s_{K_{(\cdot)}^*}(K_{(\cdot)}^t) \cdot z^t \mid \mathcal{F}_{-t}] \\
&= (K_{(\cdot)}^t + \frac{\varepsilon^t}{2}s_{K_{(\cdot)}^*}(K_{(\cdot)}^t)) \cdot \mathbb{E}[z^t \mid \mathcal{F}_{-t}] = 0
\end{aligned} \quad (11)$$

Then we take the expectation of both sides and the theorem is proved:

$$\begin{aligned}
\mathbb{E}[\|K_{(\cdot)}^* - K_{(\cdot)}^{t-1}\|^2] &= \mathbb{E}[\|K_{(\cdot)}^* - K_{(\cdot)}^t - \frac{\varepsilon^t}{2}s_{K_{(\cdot)}^*}(K_{(\cdot)}^t)\|^2] \\
&+ \varepsilon^t\mathbb{E}[\|z^t\|^2] - 2\sqrt{\varepsilon^t}\mathbb{E}[K_{(\cdot)}^* \cdot z^t]
\end{aligned} \quad (12)$$

Suppose we keep the noise variance $\mathbb{E}[\|z^t\|^2]$ unchanged, $z^t$ impacts the deviation only through the correlation term $-2\sqrt{\varepsilon^t}\mathbb{E}[K_{(\cdot)}^* \cdot z^t]$. Remarkably, since we use operators "Weight-K-Space" or "Mask-K-Space" to restrict the target data distribution and diffusion process in the high-frequency domain to explore the prior knowledge, coupled with theoretical analysis, both of them in the high-frequency diffusion process are more positive relevant. This suggests that if we have the elements of $z^t$ positively correlate with the corresponding elements of $K_{(\cdot)}^*$, this deviation will decrease, encouraging the diffusion convergence.

*C. Underlying Correlations between High-frequency Operators*

As previously discussed, "Weight-K-Space" and "Mask-K-Space" extract corresponding high-frequency prior information in k-space, so this subsection will undertake a more comprehensive exploration into the underlying elements of these two operators. Fig. 3 exhibits the Fourier-transformed images of the original k-space data after different preprocessing operators. Due to the essence of operators is to extract high-frequency information, the obtained images have the characteristic of structural similarity. Specifically, it means that the structural information garnered through the two strategies exhibits a degree of redundancy and correlation. Nevertheless, changing the kernel of "Mask-K-Space" can confirm the effect of high-frequency component injection so that the corresponding feature maps are distinct, the blue indicated line in Fig. 3 gives a comprehensive visualization. Consequently, in order to explore the correlation between the two strategies, we further employ the correlation coefficient as an evaluative measure to gauge the similarity between the feature maps associated with the "Weight-K-Space" and those relevant to the "Mask-K-Space" operators of varying sizes (depicted by the red indi-cator line in Fig. 3), i.e., $\rho(x,y) = \frac{\text{cov}(x,y)}{\sigma(x) \cdot \sigma(y)}$. where cov(•) represents the covariance between maps and $\sigma(\cdot)$ represents the corresponding standard deviation of the maps. Note that the closer the correlation coefficient is to 1, the higher the correlation between maps.

Evidently, it is apparent that the feature map denoted as $x_2$ exhibits the highest degree of correlation with the feature map $y$ originating from the "Weight-K-Space", when the kernel of operator "Mask-K-Space" is set to $50 \times 50$. It is because more high-frequency components are retained by "Mask-K-Space" with $50 \times 50$ than the small-window operator, the more perfectly the texture structure can be reconstructed. As the kernel expands to $70 \times 70$, the correlations between "Weight-K-Space" and "Mask-K-Space" exhibit a gradual reduction. One possible reason is that more high-frequency information obtained by "Mask-K-Space" is lost due to excessive pursuit of narrowing the difference. The potential phenomenon further provides additional evidence supporting the idea that the structural information extracted by the two operators shares a correlation, yet each operator also possesses its own specific focus. Ablation experiments will further confirm this phenomenon at the level of evaluation metrics.

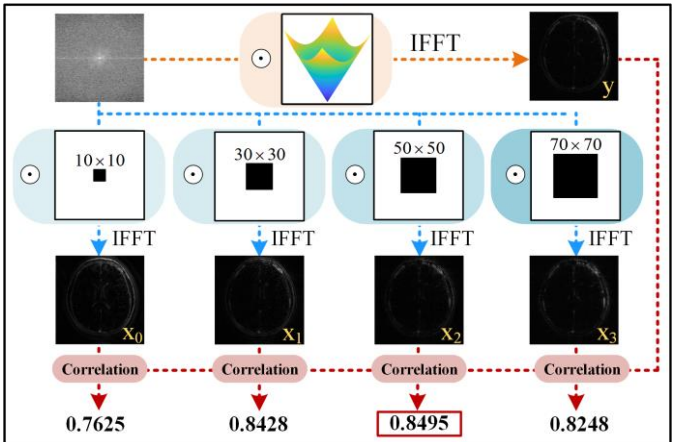

**Fig. 3.** Visualization of the underlying features of high-frequency operators. Yellow line represents underlying features in Weight-K-Space and the blue line exhibits features corresponding to different kernels of Mas-K-Space. Meanwhile, the red line shows the correlation of different feature maps.

On this basis, we combine different schemes jointly to excavate high-frequency information according to the aggregation idea to form the multi-frequency prior. By combining the advantages of distinct operators, they can be integrated with each other to obtain good interpretability of network topology and generate realistic results. Afterwards, two collaborative methodologies are introduced to capture multi-frequency prior knowledge and accelerate the diffusion process. The upper column of Fig. 1 depicts the procedure of combinational manners. One kind of combinational mode is in view of serial combination or two-stage learning, that is, "Weight-K-Space" and "Mask-K-Space" are combined in a **serial-manner**. First of all, the under-sampled data in k-space domain multiplied by the weight-based matrix is identified as the input of "Weight-K-Space". Subsequently, the output obtained by the "Weight-K-Space" operator is divided by the corresponding weight-based matrix. Meanwhile, the new input of "Mask-K-Space" emerges after being handled by the artificial mask. Another kind of combinational mode is **parallel-manner**. Note that the input of "Weight-K-Space" is multiplied by the weight-based matrix while the input of "Mask-K-Space" is the k-space data handled by the artificial mask. Finally, we conduct the mean operation on the outputs of two models.

### D. Diffusion Reconstruction via Multi-frequency Prior

For the purpose of utilizing the strong generation ability of the generated model and generating fixed data, the data consistency (DC) operation is imposed after every generation step. Thus, we can sample from the $p(k | y)$ in a reasonable way. The subproblem with regard to DC can be described as:

$$\underset{k}{Min}\{\|Ak - f\|_2^2 + \lambda \|k - K\|_2^2\} \quad (15)$$

where $k$ is the entry in k-space generated by network and $K$ stands for the under-sampled measurement in k-space domain. Due to the collaborative strategy that optimizes $K_w$ and $K_m$ recurrently by different diffusion process employed in this work, the final output needs to be weighted before data consistency, which facilitates the subsequent hybrid prior information to be fed into the network again. The $K$ can be described as the solution of the two collaborative manners:

$$K = \begin{cases} \mu_2 K_m\{(\mu_1(K_w))\}; & Serial \\ \lambda_1(K_w) + \lambda_2(K_m); & Parallel \end{cases} \quad (16)$$

where $\mu_{(\cdot)}$ and $\lambda_{(\cdot)}$ control the level of linear combination between serial and parallel values. Based on this formula, the corresponding DC solution could be solved through mathematical reasoning:

$$k(u) = \begin{cases} k(u), & if \ u \notin \Omega \\ [k(u) + \lambda K(u)]/(1+\lambda); & if \ u \in \Omega \end{cases} \quad (17)$$

where $\Omega$ denotes an index set of the acquired k-space samples. $k(u)$ is the entry at index $u$ in k-space domain generated by network. When setting noiseless (i.e., $\lambda \to \infty$), the predicted coefficient at $u$ step is substitute by the initial coefficient if it has been sampled.

Traditionally, the reconstruction problem has been conceptualized as a low-rank matrix completion problem. To achieve high-quality reconstruction results, we have incorporated traditional operator following network iteration, which serves to further restore the low-rank matrix. The object of traditional operator is a data matrix which is generated by the network. As the k-space data matrix transforms into Hankel matrix formulation, we can analyze the hard-threshold singular values of the data matrix. According to the low-rank property in Hankel matrix formulation, solving the low-rank constraint term turns to an optimization problem:

$$\underset{k}{Min} \|Ak - y\|_2^2 \ s.t. \ rank(L) = l, k = H^+(L) \quad (18)$$

where $H^+(\cdot)$ is the Hankel pseudo-inverse operator, $L$ is a data matrix with low-rank property after conducting hard-threshold singular values operation, and $l$ is the rank of the data matrix. In addition, after hard thresholding, DC operation is implemented subsequently to fix data. **Algorithm 1** explains the reconstruction algorithm in detail.

| Algorithm 1: CM-DM |
|---|
| **Required:** $S_\theta(K_w)$; $S_\theta(K_m)$ |

1: Setting: $\{K_w, K_m, z^t\} \sim N(0, I_{a \times b \times c})$ ; $1 \leq t \leq T$
2: **for** $t = T$ **to** 1 **do**
3: $\quad K_w^{t-1} = K_w^t + \frac{\varepsilon^t}{2} S_\theta(K_w) + \sqrt{\varepsilon^t} F[z^t]$
4: $\quad K_m^{t-1} = K_m^t + \frac{\varepsilon^t}{2} S_\theta(K_m) + \sqrt{\varepsilon^t} F[z^t]$
5: **end for**
6: Update $K_w(k)$ and $K_m(k)$
7: Combination: $K = \begin{cases} \mu_2 K_m\{(\mu_1(K_w))\}; & Serial \\ \lambda_1(K_w) + \lambda_2(K_m); & Parallel \end{cases}$
8: Data Consistency:
$\quad k(u) = \begin{cases} k(u), & if\ u \notin \Omega \\ [k(u) + \lambda K(u)]/(1+\lambda); & if\ u \in \Omega \end{cases}$
9: Traditional Operator:
$\quad \underset{k}{Min}\ \|Ak - y\|_2^2\ \ s.t.\ rank(L) = l, k = H^+(L)$
10: Return $k$

## IV. EXPERIMENTS

### A. Experimental Setup

In this section, the performance of CM-DM is compared with state-of-the-arts at different acceleration factors and sampling patterns. To ensure comparability and fairness of the experiments, all methods are conducted on the same datasets. Open-source code related to this study is available at: https://github.com/yqx7150/CM-DM.

*Datasets:* The brain dataset **SIAT** is provided by Shenzhen Institute of Advanced Technology, Chinese Academy of Sciences and informed consent of the imaging subject is obtained in accordance with Institutional Review Board policy. 500 complex-valued images of them are selected as the training data which are collected from healthy volunteers using a T2-weighted turbo spin echo sequence on a 3.0T scanner. These fully-sampled data are acquired by a 12-channel head coil with matrix size of $256 \times 256$ and combined to single-channel complex-valued data. Notably, the preprocessing operations enhance them to 4000 single coil images for training. To verify its generalization ability, we will test the model's reconstruction performance on in-vivo datasets with different sequences. **T1-weighted Brain** are obtained from healthy volunteers with a T1-weighted, 3D spoiled gradient echo sequence in a 1.5T MRI scanner using an 8-channel joint-only coil. The FOV is $20 \times 20 \times 20 mm^2$, TR/TE is $17.6/8ms$ and Flip angle is $20°$. **T1-GE Brain** are MR images including 8 channel complex-valued obtained using 3.0T GE. The FOV is $220 \times 220 mm^2$, and TR/TE is $700/11ms$. In addition, to perform clinical validation, we use the **FastMRI+** dataset [41] which is a large and open dataset of knee and brain MRI to validate the proposed method. Note that **Test1** and **Test2** with rich texture details in the dataset are also used to show the experimental effect.

*Parameter Configuration:* Only two different methods WKGM and CM-DM are all trained in k-space domain while the others are in the image domain. Using Adam optimizer for training with $\beta_1 = 0.9$ and $\beta_2 = 0.999$ to optimize the network. For noise variance schedule, we choose $\sigma_{max} = 1$, $\sigma_{min} = 0.01$ and $r = 0.075$. Besides, we use $N = 1000$ and $M = 1$ iterations for inference as default, unless specified otherwise. For other parameters, we empirically tune the parameters in their suggested ranges to give their best performances. The training and testing experiments are performed with 2 NVIDIA TITAN GPUs, 12 GB.

*Performance Evaluation:* To quantitatively measure the error caused by CM-DM, the Peak Signal-to-Noise Ratio (PSNR), Structural Similarity (SSIM), and Mean Squared Error (MSE) are used to evaluate the quality of reconstruction images. Note that quantitative evaluations are all computed on the image domain. For ease of calculation, reconstructed and reference images are derived using an inverse Fourier transform followed by an elementwise sum of squares (SOS) operation.

TABLE I
PSNR, SSIM, AND MSE (*E$^{-4}$) COMPARISON WITH STATE-OF-THE-ART METHODS UNDER POISSON, 2D RANDOM, AND UNIFORM SAMPLING PATTERNS WITH VARYING ACCELERATION FACTORS.

| *T1-GE Brain* | SAKE | P-LORAKS | EBMRec | HGGDP | **CM-DM** |
|---|---|---|---|---|---|
| Poisson $R$=10 | 38.33/0.9207/1.470 | 36.05/0.8723/2.484 | 29.60/0.7263/10.970 | 32.80/0.8986/5.246 | **40.58/0.9387/0.901** |
| Poisson $R$=15 | 34.99/0.8975/3.168 | 32.83/0.7996/5.211 | 26.90/0.6876/20.397 | 30.94/0.8685/8.049 | **38.50/0.9264/1.413** |
| 2D Random $R$=8 | 28.40/0.8322/14.447 | 34.07/0.8226/3.915 | 25.76/0.6508/26.561 | 31.46/0.8541/7.140 | **39.28/0.9337/1.181** |
| 2D Random $R$=12 | 25.59/0.7733/27.574 | 30.29/0.7318/9.349 | 26.12/0.5829/24.456 | 29.06/0.8038/12.429 | **37.71/0.9142/1.693** |
| *T1-weighted Brain* | SAKE | P-LORAKS | EBMRec | HGGDP | **CM-DM** |
| Uniform $R$=8 | 27.78/0.7348/16.688 | 28.15/0.7175/15.311 | 28.12/0.6805/15.429 | 29.06/0.7980/12.428 | **36.55/0.8899/2.215** |
| Uniform $R$=10 | 26.62/0.7058/21.768 | 28.06/0.7172/15.640 | 27.36/0.6407/18.366 | 28.06/0.7872/15.618 | **35.54/0.8831/2.791** |
| 2D Random $R$=12 | 27.09/0.6460/19.546 | 30.29/0.7019/9.362 | 29.06/0.6689/12.413 | 29.85/0.7496/10.349 | **36.01/0.8527/2.504** |
| 2D Random $R$=15 | 26.97/0.6360/20.082 | 28.89/0.6750/12.912 | 28.45/0.6701/14.284 | 29.40/0.7407/11.475 | **35.34/0.8513/2.924** |

### B. Reconstruction Experiments

**Comparisons with State-of-the-arts.** To verify the advantages of the proposed method, we conduct a comparative experiment with the following DL-based methods (HGGDP [37] and EBMRec [38]) and traditional methods (SAKE [39] and P-LORAKS [40]). Meanwhile, it is mentioned that algorithm CM-DM has opted for the serial-manner in subsequent experiments. The quantitative results of the above methods evaluated under 8×, 10×, 12×, and 15× acceleration factors for Poisson, random, and uniform sampling patterns are shown in Table I. As can be seen from Table I, CM-DM performs best comparing to others for all acceleration factors and all sampling patterns. More particularly, as the ac-celeration factor increases, the performance of CM-DM is still considerable. Overall, CM-DM achieves the best performance in terms of the quantitative metrics and preserves the most realistic high-frequency details.

To further prove the superiority of CM-DM over other algorithms, the visualization results of different methods under different acceleration factors are shown in Figs. 4-5. Overall, as for processing high-frequency data, CM-DM achieves more accurate reconstruction with clear texture and boundaries. For example, compared to the reference image, SAKE shows significant noise-like residuals associated with high acceleration noise vulnerability. More obviously, P-LORAKS shows blurred reconstruction details, which is effectively alleviated by CM-DM. Theoretically, the inferiority of P-LORAKS and SAKE indicates that the traditional model is not robust enough while meeting challenging data with amounts of high-resolution features and details. As one of the generation models, EBMRec suffers from heavy noise and artifacts under high acceleration factors. Although HGGDP makes some progress compared with EBMRec, it is not devoted to high-frequency prior information and obtains unsatisfactory effects under high acceleration factors. The above experimental results validate that CM-DM can accurately reconstruct images and has good generalization ability.

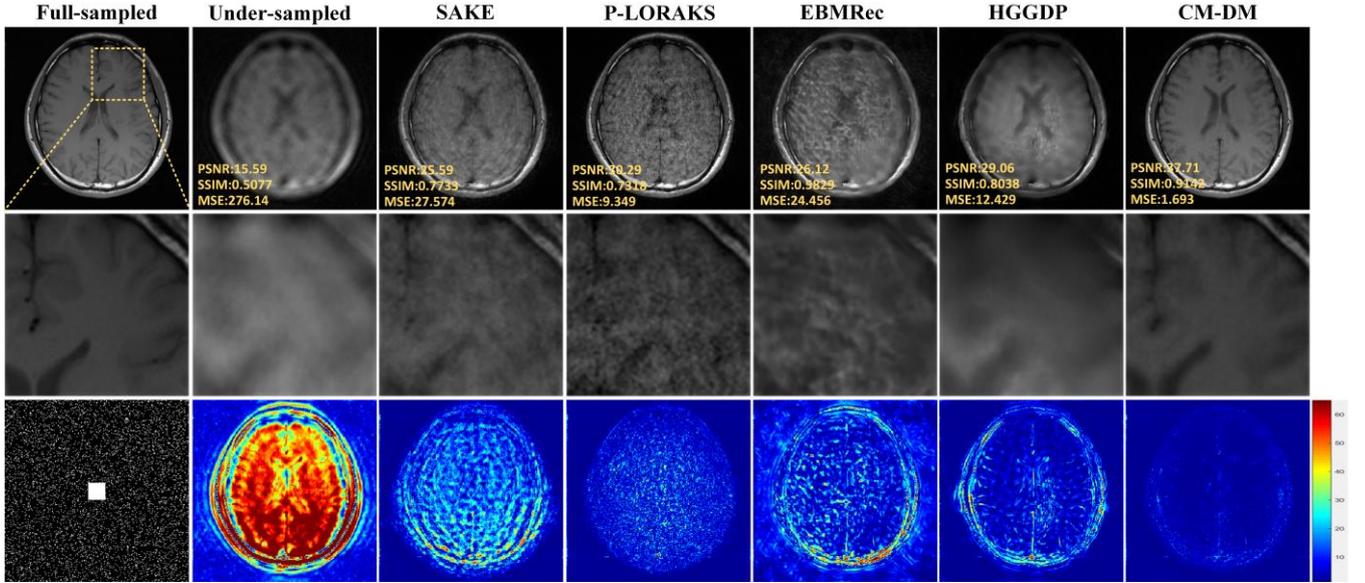

**Fig. 4.** Reconstruction of the ***T1-weighted Brain*** at random sampling of *R*=12. From left to right: Full-sampled, Under-sampled, reconstruction by SAKE, P-LORAKS, EBMRec, HGGDP, and CM-DM (ours). The second row shows the enlarged view of the ROI region (indicated by the yellow box in the first row), and the third row shows the error map of the reconstruction.

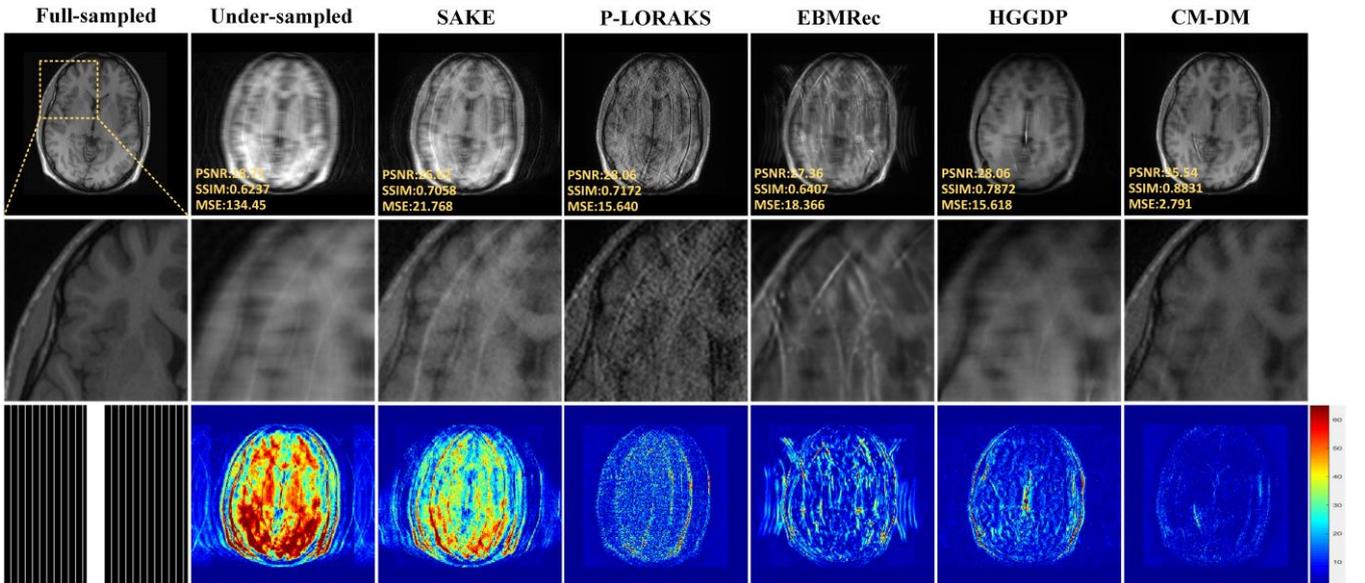

**Fig. 5.** Complex-valued reconstruction results at *R*=10 using uniform sampling with 8 coils. From left to right: Full-sampled, Under-sampled, reconstruction by SAKE, P-LORAKS, EBMRec, HGGDP, and CM-DM (ours). The second row shows the enlarged view of the ROI region (indicated by the yellow box in the first row), and the third row shows the error map of the reconstruction.

***Diffusion Models in High-frequency Domain.*** We also designed comparison experiments with the score-based diffusion methods WKGM [22] and HFS-SDE [23] trained in high-frequency domain to evaluate the reconstruction accuracy of CM-DM. The quantitative metrics for different acceleration factors and various sampling patterns are provided in Table II. Especially, when subjected to the 2D random sampling pattern at an acceleration factor of *R*=15, CM-DM exhibits superior performance compared to WKGM by a margin of 4.39 dB in terms of PSNR. The phenomenon in-dicates that the combination of multiple operators for extracting high-frequency priors can promote the generation of more details in the reconstruction process.

Furthermore, Fig. 6 visually highlight the strengths of CM-DM. CM-DM owns lower residual errors and the higher sensitivity when describing detailed tissue structures. It means that CM-DM retrieves more high-frequency information. In view of the fact that finer details of the image are associated with the high-frequency components, our proposed method demonstrates the capability to retain well-defined boundaries and textural intricacies, resulting in a precise representation of the reconstructed image. Besides, distinct from the existing methods WKGM and HFS-SDE that employ a solitary high-frequency prior, CM-DM utilizes different strategies to combine multi-frequency priors for achieving superior accuracy in MRI reconstruction.

TABLE II
PSNR, SSIM, AND MSE (*E$^{-4}$) COMPARISON WITH DIFFUSION MODELS OF HIGH-FREQUENCY DOMAIN UNDER 2D RANDOM AND CARTESIAN SAMPLING PATTERNS WITH VARYING ACCELERATE FACTORS.

| *T1-GE Brain* | HFS-SDE | WKGM | CM-DM |
|---|---|---|---|
| 2D Random $R$=10 | 36.19 | 33.76 | **38.30** |
|  | 0.8324 | 0.9039 | **0.9200** |
|  | 2.403 | 4.203 | **1.479** |
| 2D Random $R$=15 | 33.58 | 31.94 | **36.33** |
|  | 0.7836 | 0.8794 | **0.9083** |
|  | 4.384 | 6.395 | **2.326** |
| *Test 1* | HFS-SDE | WKGM | CM-DM |
| Cartesian $R$=10 | 34.88 | 34.93 | **35.40** |
|  | 0.8753 | 0.8805 | **0.9070** |
|  | 3.245 | 3.210 | **2.883** |
| Cartesian $R$=15 | 33.52 | 33.63 | **34.04** |
|  | 0.8119 | 0.8586 | **0.8905** |
|  | 4.445 | 4.335 | **3.940** |

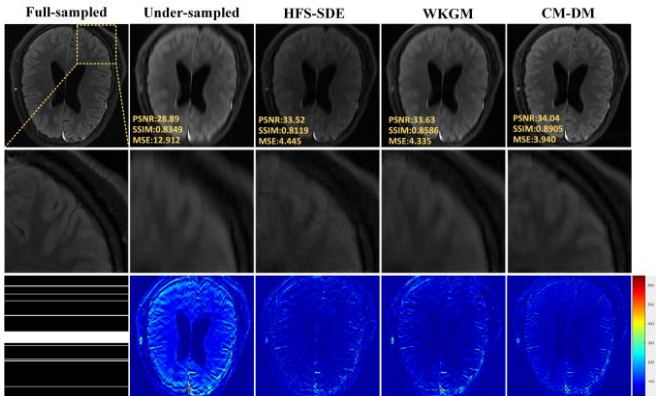

**Fig. 6.** Complex-valued reconstruction results at $R$=15 using Cartesian sampling with 12 coils. From left to right: Full-sampled, Under-sampled, reconstruction by HFS-SDE, WKGM, and CM-DM (ours). The second row shows the enlarged view of the ROI region (indicated by the yellow box in the first row), and the third row shows the error map of the reconstruction.

### C. Performance of Preserving Pathological Regions

Experiments are conducted with E2E-Varnet [42] via the ***Test 2*** to evaluate the clinical feasibility of CM-DM. It can be seen that the proposed method is able to reconstruct the structural content in the image, including many fine details, successfully. This is also indicated by the quantitative results shown in Table III that CM-DM outperforms E2E-Varnet by a large margin in all cases. As the acceleration factor is increased, we see that the uncertainty increases correspondingly. Therefore, although the pathological area is enlarged to clarify the details, some noise still remains in the results of CM-DM. By automatically mapping the intrinsic feature from the MR images, E2E-Varnet can also provide comparative results for MRI reconstruction, which is consistent with the observations in Fig. 7. Nevertheless, the difference with the full-sampled image indicates that it is too smooth to fully preserve the structural content. Although the SSIM value of E2E-Varnet and CM-DM are similar, over smooth and distort phenomena can be easily recognizable while accurate texture details with few noises are reconstructed by CM-DM. Obviously, CM-DM shows successful effects in maintaining pathological regions of the image.

TABLE III
PSNR, SSIM, AND MSE (*E$^{-4}$) MEASURES FOR DIFFERENT ALGORITHMS WITH VARYING ACCELERATE FACTORS.

| *Test 2* | Zero-filled | E2E-Varnet | CM-DM |
|---|---|---|---|
| Equispaced $R$=10 | 19.30 | 25.61 | **28.27** |
|  | 0.8207 | 0.8987 | **0.8995** |
|  | 117.531 | 27.507 | **14.877** |
| Equispaced $R$=12 | 17.846 | 24.43 | **26.06** |
|  | 0.7899 | 0.8828 | **0.8836** |
|  | 126.849 | 36.025 | **24.770** |
| Equispaced $R$=15 | 17.123 | 22.56 | **24.03** |
|  | 0.7727 | 0.8525 | **0.8652** |
|  | 135.608 | 46.636 | **39.537** |

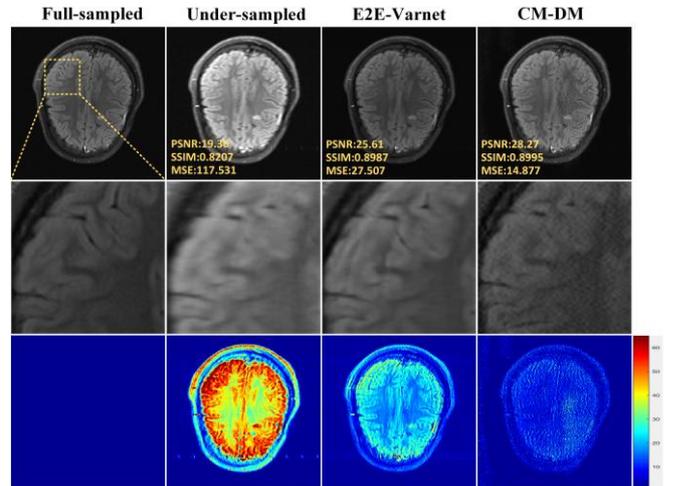

**Fig. 7.** Reconstruction results using E2E-Varnet and CM-DM at $R$=10 of the equispaced mask. From top to bottom: Reconstruction images, magnified views corresponding to pathological regions, residual maps.

### D. Convergence Analysis

In this subsection, the correlation between the convergence of WKGM and CM-DM with the number of iterations is investigated by quantitative indices PSNR and SSIM. We randomly select an example of reconstructing brain images using the random sampling pattern with an acceleration factor $R$=8. It can be seen in Fig. 8, both the SSIM curve and the PSNR curve of WKGM model and CM-DM model first rise rapidly with the number of iterations increasing then gradually stabilize. It means that the reconstruction method in high-frequency domain is prone to converge due to the small range of high-frequency information, which is also proved mathematically in **Theorem 1**. Therefore, it can be concluded that the reconstruction method directly constraining the model in k-space leads to fewer iterations. However, the PSNR and SSIM curves of CM-DM reach convergence in a quicker pace in Fig. 8. It is encouraged that the combination of two different high-frequency distributions which is correlated and more random can increase the similarity between the sample initialization and noise sampling. Consequently, the scheme of model combination is beneficial to capture sufficient high-frequency information and accelerate convergence.

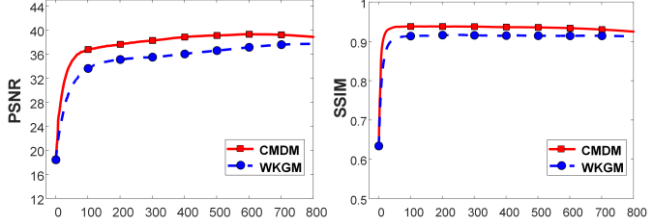

Fig. 8. Convergence curves of WKGM and CM-DM in terms of PSNR and SSIM versus the iteration number when reconstructing the brain image from 1/8 sampled data under random sampling pattern.

## V. DISCUSSION

We have demonstrated that spatially constraining the diffusion process in high-frequency domain effectively improves reconstruction stability and convergence speed. Moreover, CM-DM combines different high-frequency distributions to increase the correlation of the data, thus reducing the time required for sampling while maintaining the reconstruction quality. However, some areas still need further discussion or improvement for our proposed model.

On the one hand, we extract multi-frequency priors with various combination manners and make a quick comparison in the same situation. The collaborative strategies include the serial-manner and the parallel-manner. Here, we choose the serial-manner as an example. Detailed experimental results are demonstrated in the Table IV. Similarly, "W" corresponds to the "Weight-K-Space". It is evident that the performance of two model combination always outperforms than that of the single model. Meanwhile, the combination of the same type model slightly improves the quantitative effects. The combination of "Weight-K-Space" and "Mask-K-Space" makes a big difference. When choosing the combination of the weighting scheme and the "Mask-K-Space" size $50 \times 50$, every index scores the best. As a result, this situation is picked as CM-DM.

TABLE IV
COMPARISON OF PSNR, SSIM, AND MSE (*E$^{-4}$) OF MODEL COMBINATION MODES IN POISSON AND 2D RANDOM SAMPLING PATTERNS.

| T1-GE Brain | Poisson R=10 | 2D Random R=10 |
|---|---|---|
| W1-W2 | 38.51/0.9492/1.409 | 33.33/0.8984/4.650 |
| M30-M50 | 39.53/0.9197/1.115 | 35.43/0.8692/2.863 |
| W-M30 | 40.44/0.9364/0.903 | 37.09/0.9060/1.952 |
| W-M50 | **40.58/0.9387/0.901** | **38.30/0.9200/1.479** |
| W-M70 | 41.35/0.9421/0.733 | 38.07/0.9180/1.557 |

On the other hand, exploration of diverse combination manners to attain multi-frequency prior is imperative owing to the variety of high-frequency operators. In light of the previous discussion, we executed a set of ablation experiments. Given that the amalgamation of two preconditioning operators enables the extraction of substantial prior information within the multi-frequency domain, both parallel and serial combinations manners achieve performance gains. Notably, quantitative analysis of the results in Table V also indicates that the gains obtained by different combination manners are slightly different from the practical point of view. Generally speaking, the reconstruction effect of the series-manner is more dominant and it has been selected in this study.

TABLE V
COMPARISON OF PSNR AND SSIM OF DIFFERENT COMBINATION MANNERS IN 2D RANDOM SAMPLING PATTERNS.

| T1-GE Brain | Manners | WKGM | CM-DM |
|---|---|---|---|
| 2D Random R=6 | Parallel | 39.94/0.960 | 41.90/0.955 |
|  | Serial | 39.94/0.960 | **41.99/0.957** |
| 2D Random R=10 | Parallel | 33.76/0.904 | 35.44/0.888 |
|  | Serial | 33.76/0.904 | **38.29/0.916** |

Due to the large iteration numbers and long reconstruction time of the diffusion model, improving the sampling speed and the generating speed of the diffusion model remains an ongoing study. Although CM-DM achieves accurate reconstruction while in need of relatively short time, there is still large room for shortening the reconstruction time. Therefore, how to minimize the generation time when reconstructing accurate images is the future goal.

## VI. CONCLUSION

In conclusion, we present a novel mathematical framework with correlated and multi-frequency prior to preserve the structural details in the reconstructed image. Specifically, different preprocessing operators are designed to solve the intractable problem of large magnitude contrast between low-frequency and high-frequency k-space data. Meanwhile, we reformulate the diffusion process with corelated and multi-frequency prior information whilst preserving its steady-state distribution, so that the noise and target distributions are much closer to dramatically speed up the convergence of sampling process. Further theoretical analyses and thorough experimental validations fully validate that CM-DM outperforms existing MRI reconstruction methods and achieves comparable performances with the conventional score-based diffusion methods. Other than that, there still remain unanswered questions. Hence, we expect that many interesting questions and answers will be actively discussed in the near future.

## APPENDIX

### A. Diffusion Process in Frequency Domain

In this section, we will prove theoretically why we can directly regulate the frequency distribution of a diffusion process through the preconditioning strategy, and why it is necessary to do so. We first show that the sampling process can be equivalently transformed to the frequency domain via an orthogonal transform. We start with the following lemma.

**Lemma A. 1.** If two $d$-dimensional random vectors $x$, $y \in \mathbb{R}^d$ have differentiable density functions, and satisfy $y = Gx$, where the matrix $G \in \mathbb{R}^{d \times d}$ is invertible, we have

$$\nabla_y \log p_y(y) = \nabla_y \log p_x(G^{-1}y) = \nabla_y \log p_x(x) \quad \text{(A-1)}$$

**Proof.** Note for any invertible differentiable transformation $g \in \mathbb{R}^d \to \mathbb{R}^d$, if $y = g(x)$, we have:

$$p_y(y) = p_x(g^{-1}(y)) \left| \det\left[ \frac{dg^{-1}(y)}{dy} \right] \right| \quad \text{(A-2)}$$

In particular, $p_y(y) = p_x(G^{-1}y)|G|^{-1}$. We verify the lemma by taking the logarithm and calculating the gradients at both sides of the equation.

**Theorem A. 1.** Suppose $F \in \mathbb{R}^{d \times d}$ is an orthogonal matrix, then the diffusion process (Eq. (1)) can be rewritten as

$$k^{t-1} = k^t + \frac{\varepsilon^t}{2}\nabla_k \log p_k(k^t) + \sqrt{\varepsilon^t} F[z^t] \quad \text{(A-3)}$$

where $k = Fx$, and $\{z^t\}$ do not need to follow isotropic Gaussian distributions.

*Proof.* Multiplying $F$ at both sides of the original diffusion process, we have

$$k^{t-1} = k^t + \frac{\varepsilon^t}{2}F\nabla_x \log p_x(x^t) + \sqrt{\varepsilon^t} F[z^t] \quad \text{(A-4)}$$

By Lemma A. 1, we have $\nabla_k \log p_k(k) = \nabla_k \log p_x(x)$. We also have $\nabla_x = F^T \nabla_k$ by the chain rule. Putting all the things together, we have

$$k^{t-1} = k^t + \frac{\varepsilon^t}{2}FF^T \nabla_k \log p_k(k^t) + \sqrt{\varepsilon^t} F[z^t] \quad \text{(A-5)}$$

As $F$ is orthogonal, there exists $FF^T = I_d$. We thus finish the proof.

## REFERENCE

[1] A. J. Sederman, L. F. Gladden, and M. D. Mantle, "Application of magnetic resonance imaging techniques to particulate systems," *Adv. Powder Techn.*, vol. 18, pp. 23–38, 2007.

[2] R. Stannarius, "Magnetic resonance imaging of granular materials," *Rev. Sci. Instr.*, vol. 88, no. 5, 2017.

[3] M. Lustig, D. Donoho, and J. M. Pauly, "Sparse MRI: The application of compressed sensing for rapid MR imaging," *Magn. Reson. Med.*, vol. 58, no. 6, pp. 1182-1195, 2007.

[4] U. Gamper, P. Boesiger, and S. Kozerke, "Compressed sensing in dynamic MRI," *Magn. Reson. Med.*, vol. 59, no. 2, pp. 365-373, 2008.

[5] H. Jung, K. Sung, K. S. Nayak, E. Y. Kim, and J. C. Ye, "k-t FOCUSS: a general compressed sensing framework for high resolution dynamic MRI," *Magn. Reson. Med.*, vol. 61, no. 1, pp. 103-116, 2009.

[6] D. L. Donoho, "Compressed sensing," *IEEE Trans. Inform. Theory*, vol. 52, no. 4, pp. 1289-1306, 2006.

[7] D. K. Sodickson, and W. J. Manning. "Simultaneous acquisition of spatial harmonics (SMASH): fast imaging with radiofrequency coil arrays," *Magn. Reson. Med.*, vol. 38, no. 4, pp. 591-603, 1997.

[8] K. P. Pruessmann, M. Weiger, M. B. Scheidegger, and P. Boesiger, "SENSE: sensitivity encoding for fast MRI," *Magn. Reson. Med.*, vol. 42, no. 5, pp. 952-962, 1999.

[9] M. A. Griswold, P. M. Jakob, R. M. Heidemann, M. Nittka, V. Jellus, J. Wang, B. Kiefer, and A. Haase, "Generalized autocalibrating partially parallel acquisitions (GRAPPA)," *Magn. Reson. Med.*, vol. 47, no. 6, pp. 1202–1210, 2002.

[10] Y. Song, C. Durkan, I. Murray, et al., "Maximum likelihood training of score-based diffusion models," *Adv. neural inf. process. syst.,* vol. 34, pp. 1415-1428, 2021.

[11] D. Kwon, Y. Fan, and K. Lee, "Score-based generative modeling secretly minimizes the Wasserstein distance," *Adv. neural inf. process. syst.*, vol. 35, pp. 20205-20217, 2022.

[12] S. Ghimire, J. Liu, A. Comas, D. Hill, et al., "Geometry of score based generative models," *arXiv preprint arXiv:* 2302.04411, 2023.

[13] Y. Song and S. Ermon, "Generative modeling by estimating gradients of the data distribution," *Adv. neural inf. process. syst.*, vol. 32, 2019.

[14] J. Ho, A. Jain, and P. Abbeel, "Denoising diffusion probabilistic models," *Adv. neural inf. process. syst.*, vol. 33, pp. 6840-6851, 2020.

[15] J. Song, C. Meng, and S. Ermon, "Denoising diffusion implicit models," *arXiv preprint arXiv*: 2010.02502, 2020.

[16] C. Quan, J. Zhou, Y. Zhu, Y. Chen, S. Wang, D. Liang, and Q. Liu, "Homotopic gradients of generative density priors for MR image reconstruction," *IEEE Trans. Med. Imag.*, vol. 40, no. 12, pp. 3265-3278, 2021.

[17] H. Chung, and J. C. Ye, "Score-based diffusion models for accelerated MRI," *Med. Image. Anal.*, vol. 80, p. 102479, 2022.

[18] P. Deora, B. Vasudeva, S. Bhattacharya, and P. M. Pradhan, "Structure preserving compressive sensing MRI reconstruction using generative adversarial networks," *Proc. IEEE Conf. Comput. Vis. Pattern Recognit.*, pp. 522-523, 2020.

[19] F. A. Razzaq, S. Mohamed, A. Bhatti, and S. Nahavandi, "Non-uniform sparsity in rapid compressive sensing MRI," IEEE Trans. Syst. Man Cybern. Syst., pp. 2253-2258, 2012.

[20] K. Sung and B. A. Hargreaves. "High-frequency subband compressed sensing MRI using quadruplet sampling," *Magn. Reson. Med.*, vol. 70, no. 5, pp. 1306-1318, 2012.

[21] Y. Xie, and Q. Li, "Measurement-conditioned denoising diffusion probabilistic model for under-sampled medical image reconstruction," *Proc. Int. Conf. Med. Image Comput. Comput.-Assisted Intervention*, pp. 655-664, 2022.

[22] Z. Tu, D. Liu, X. Wang, C. Jiang, M. Zhang, Q. Liu, and D. Liang, "WKGM: Weight-k-space generative model for parallel imaging reconstruction," *arXiv preprint arXiv: 2205.03883*, 2022.

[23] C. Cao, Z. X. Cui, S. Liu, D. Liang, and Y. Zhu, "High-frequency space diffusion models for accelerated MRI," arXiv preprint arXiv: 2208.05481, 2022.

[24] H. Chung, B. Sim, and J. C. Ye, "Come-closer-diffuse-faster: Accelerating conditional diffusion models for inverse problems through stochastic contraction," *Proc. IEEE Conf. Comput. Vis. Pattern Recognit.*, pp. 12413-12422, 2022.

[25] Y. Song, and S. Ermon, "Improved techniques for training score-based generative models," *Adv. neural inf. process. syst.*, vol. 33, pp. 12438-12448, 2020.

[26] T. Dockhorn, A. Vahdat, and K. Kreis, "Score-based generative modeling with critically-damped langevin diffusion," *arXiv preprint arXiv:* 2112.07068, 2021.

[27] A. Jalal, M. Arvinte, G. Daras, E. Price, A. G. Dimakis, J. Tamir, "Robust compressed sensing MRI with deep generative priors," *Adv. Neural Inf. Process. Syst.*, vol. 34, pp. 14938–14954, 2021.

[28] Y. Song, L. Shen, L. Xing, and S. Ermon, "Solving inverse problems in medical imaging with score-based generative models," *arXiv preprint arXiv:* 2111.08005, 2021

[29] A. Gungor, S. U. Dar, S. Ozturk, Y. Korkmaz, H. A. Bedel, et al., "Adaptive diffusion priors for accelerated MRI reconstruction," *Med. Imag. Anal.*, 102872, 2023.

[30] K. Sung, and B. A. Hargreaves, "High-frequency sub-band compressed sensing MRI using quadruplet sampling," *Magn. Reason. Med.*, vol. 70, no. 5, pp. 1306-1318, 2013.

[31] Z. He, K. Hong, J. Zhou, D. Liang, Y. Wang, and Q. Liu, "Deep frequency-recurrent priors for inverse imaging reconstruction," *Signal Process.*, vol. 190, p. 108320, 2022.

[32] Y. Yang, F. Liu, W. Xu, and S. Crozier, "Compressed sensing MRI via two-stage reconstruction," *IEEE Trans. Biomed. Engineer.*, vol. 62, no.1, pp. 110-118, 2014.

[33] M. Welling, and Y. W. Teh, "Bayesian learning via stochastic gradient Langevin dynamics," *Proc. Int. Conf. Mach. Learn.*, pp. 681-688, 2011.

[34] H. Ma, L. Zhang, X. Zhu, J. Zhang, and J. Feng, "Accelerating score-based generative models for high-resolution image synthesis," *arXiv preprint arXiv:*2206.04029, 2022.

[35] V. A. Van der Schaaf, and J. V. van Hateren, "Modelling the power spectra of natural images: statistics and information," *Vision. Res.*, vol. 36, no. 17, pp. 2759-2770, 1996.

[36] Y. Song, J. Sohl-Dickstein, D. P. Kingm, A. Kumar, S. Ermon, and B. Poole, "Score-based generative modeling through stochastic differential equations," *arXiv preprint arXiv:* 2011.13456, 2020.

[37] K. Dabov, A. Foi, V. Katkovnik, et al., "Image denoising by sparse 3-D transform-domain collaborative filtering," *IEEE Trans. Image Process.*, vol. 16, no. 8, pp. 2080-2095, 2007.

[38] Y. Guan, Z. Tu, S. Wang, et al., "MRI reconstruction using deep energy-based model," *NMR Biomed.*, pp. 1-19, 2022.

[39] P. J. Shin, P. E. Larson, M. A. Ohliger, M. Elad, J. M. Pauly, D. B. Vigneron, and M. Lustig, "Calibrationless parallel imaging reconstruction based on structured low-rank matrix completion," *Magn. Reason. Med.*, vol. 72, no. 4, pp. 959-970, 2014.

[40] J. P. Haldar, and J. Zhuo, "P-LORAKS: Low-rank modeling of local k-space neighborhoods with parallel imaging data," *Magn. Reason. Med.*, vol. 75, no. 4, pp. 1499-1514, 2016.

[41] J. Zbontar, F. Knoll, A. Sriram, *et al.* "fastMRI: An open dataset and benchmarks for accelerated MRI," *arXiv preprint arXiv:* 1811.08839, 2018.

[42] A. Sriram, J. Zbontar, T. Murrell, *et al.*, "End-to-end variational networks for accelerated MRI reconstruction," *Proc. Int. Conf. Med. Image Comput. Comput.-Assisted Intervention.*, pp. 64-73, 2020.